# Current – voltage characteristics in magnetic field for layered superconductor with intrinsic pinning of Josephson vortices

T.B. Charikova[1], V.N. Neverov[1], M.R. Popov[1], S.D. Popov[1]*, N.G. Shelushinina[1], A.A. Ivanov[2]

[1]M.N. Mikheev Institute of Metal Physics, Ural Branch, Russian Academy of Sciences, 18, S. Kovalevskoy St., Ekaterinburg, 620077, Russia

[2]National Research Nuclear University MEPhI, Moscow, 115409, Russia

*e-mail: *staspopov2000@gmail.com*

**Statements and declarations**

No funds, grants, or other support was received.

**Abstract**

The results of measurements of the transverse voltage in the Hall configuration, $U_y$, under conditions of motion of the charge and vortex subsystems in orthogonal magnetic and electric fields on epitaxial films of the layered high-temperature superconductor $Nd_{2-x}Ce_xCuO_4$ are presented. The motion of Josephson vortices across $CuO_2$ layers (along the $c$-axis) is analyzed considering the effects of intrinsic pinning. An exponential dependence of critical depinning current $j_c(B)$ for Josephson vortices is established in contrast to the power law dependences for Abrikosov vortices. With increasing external magnetic field after the depinning by a magnetic field at $B$=$B_c$, the onset of the vortex creep regime, which evolves into the flux-flow regime at crossover field $B$=$B_f$, is observed.

## 1. Introduction

Under conditions of the non-stationary Josephson effect, when oscillating parameters of the Josephson junction arise at a non-zero potential difference across the barrier, there is motion of vortex structures in the mixed state of anisotropic layered superconductors. As in the case of the stationary Josephson effect, the behavior of a long Josephson junction is similar to the behavior of type-II superconductors in a magnetic field parallel to the junction plane, with the formation of single Josephson vortices in weak magnetic fields $B \geq B_{c1}$ and a subsequent transition to a dynamic mixed state corresponding to a moving vortex lattice [1].

High-temperature superconducting (HTSC) cuprate compounds have a well-defined layered structure formed by $CuO_2$ planes. In addition to traditional pinning by defects, layered superconductors employ another pinning mechanism for vortex motion perpendicular to the layers. This intrinsic pinning occurs due to the energy change when a vortex crosses a layer. When the magnetic field, $\boldsymbol{B}$, and transport current, $\boldsymbol{j}$, are perpendicular to each other and are lying down in the plane of the $CuO_2$ layers (the basal plane of the crystal), this intrinsic interlayer pinning can be dominant in samples with low defect concentrations.

In this work, we investigated the current-voltage characteristics of $Nd_{2-x}Ce_xCuO_4/SrTiO_3$ films, taken in a constant magnetic field from Hall contacts in the configuration $\boldsymbol{j}||CuO_2$, $\boldsymbol{B}||CuO_2$ and $\boldsymbol{j}\perp\boldsymbol{B}$, that is, we investigated the current-voltage characteristics $U_y(j_{ab-plane})$.

Latyshev and Volkov [2], as well as Jurgens et al. [3] studied the current-voltage characteristics of single-crystal samples of $Bi_2Sr_2CaCu_2O_x$ [2] or mesa structures based on them [3] in a magnetic field, taking the measurements of voltage from potential contacts in the configuration $\boldsymbol{j}\perp CuO_2$, $\boldsymbol{B}||CuO_2$ and $\boldsymbol{j}\perp\boldsymbol{B}$**,** that is, they investigated the dependences $U_x(j_{c-axis})$.

Despite the difference in geometric configurations, the experimental results on the evolution of the *I–V* characteristics in a magnetic field for the two studied systems are in many ways similar, since the main role in both cases is played by the generation of a flow of Josephson vortices (flux-flow regime) when the magnetic field is turned on.

The difference is the manifestation of intrinsic pinning effects when Josephson vortices move across the $CuO_2$ layers (along the *c* - axis) in our situation, and the absence of such effects in [2.3] when the flow of vortices goes along the $CuO_2$ planes.

Using the example of the $Nd_{2-x}Ce_xCuO_4$ compound and the fabricated $Nd_{2-x}Ce_xCuO_4/SrTiO_3$ epitaxial films, we investigated the transverse voltage in the Hall configuration during the motion of the charge and vortex subsystems in crossed magnetic and electric fields.

## 2. Theoretical background

Investigating the Hall effect in a mixed state provides important information about the vortex dynamics of high-temperature superconductors. Magnetic field *B* penetrates a type II superconductor via quantum vortices [4,5]. Vortex motion along the Lorentz force (perpendicular to the transport current $\boldsymbol{j}$) creates a dissipative field ($\boldsymbol{E} \parallel \boldsymbol{j}$) and leads to the flux-flow resistance. Conversely, vortex motion along the transport current direction leads to the generation of a Hall electric field ($\boldsymbol{E_H} \perp \boldsymbol{j}, \boldsymbol{B}$).

The essential difference between anisotropic layered superconductors and homogeneous isotropic ones is the presence of a strong interaction between the vortices and the crystal structure itself, known as “intrinsic pinning”: the energy of the vortex becomes dependent on the position of the vortex relative to the planes and creates a pinning force that tries to hold the vortices between the superconducting layers.

In a magnetic field parallel to the conducting $CuO_2$ planes, the vortices look the same as in Josephson junctions in superconductor/insulator/superconductor structures, since the layered structure of HTSC forms intrinsic Josephson junctions [6,7]. Josephson vortices, each of which carries a quantum of magnetic flux and whose center is located between the superconducting layers, do not have a normal core and therefore do not strongly suppress the order parameter in adjacent superconducting planes [8]. Josephson vortices can easily move along the superconducting layers, but not perpendicular to them [7]. For long Josephson junctions in a parallel magnetic field, there is a certain magnetic field at which the vortices begin to overlap, forming a triangular lattice of Josephson vortices [8]. The motion of the Josephson lattice under the action of the transport current leads to the appearance of flux-flow resistance.

The theory of intrinsic pinning for Josephson vortices in a layered superconductor was developed in [9.10.11] in a weakly layered (nearly continuous) limit [4]. Among many issues, the situation was considered in which the magnetic field $\boldsymbol{B}$, and hence the vortices, are directed parallel to the layers. The transport current $\boldsymbol{j}$ flows in the plane of the layers such that the Lorentz force is perpendicular to the layers and directed along the crystallographic *c*-direction. The displacement of the vortex along the *c*-axis is counteracted by the intrinsic pinning force. By balancing the Lorentz force and the pinning force, the magnitude of the depinning current $j_c$ at a given $B$ or critical depinning magnetic field $B_c$. at a given current $j$ can be determined.

In this work, we will consider the phenomenon of strong intrinsic pinning of Josephson vortices in the studied $Nd_{2-x}Ce_xCuO_4$.structure. It is known that intrinsic pinning in cuprate perovskite-like superconductors is a consequence of the layered structure of the material, with superconductivity occurring in the $CuO_2$ layers, while the intermediate buffer layers act as charge reservoirs (see the detailed review by Blatter et al. [12], as well as the monographs by Kopnin [4] and Tinkham [5]). Therefore, the superconducting order parameter (and, along with it, the condensation energy) is expected to exhibit strong oscillations with a period $d$, where $d$ is the distance between the $CuO_2$ planes.

The first quantitative analysis of intrinsic pinning in layered superconductors was performed by Tachiki and Takahashi [13]. They suggested that modulation of the order parameter perpendicular to the layers can pin vortices between the layers, which leads to the intrinsic pinning mechanism of Josephson vortices when magnetic field $B$ is parallel to the $CuO_2$ planes. In this case, a vortex lattice is formed in the layered structure such that the vortex centers are located between the superconducting $CuO_2$ planes.

The Lorentz force, generated by the magnetic field acting on charges moving along these planes (under the influence of an external electric field) and directed along the *c*-axis, will act on the vortices, setting them in motion. As they move along the *c*-axis, the vortices must intersect the superconducting layers, which entails a significant expenditure of condensation energy, thereby creating internal barriers—intrinsic pinning.

It should be noted that when the electric current ***j*** is directed along the *c*-axis, the Josephson vortices will shift toward the $CuO_2$ plane, where there are no obstructions to the flow. In this case, point defects are necessary to create a finite critical current density.

Intrinsic pinning was observed in $YBa_2Cu_3O_{7-\delta}$ single crystal with a layered structure in crossed magnetic fields to obtain the angular dependence of the flux-flow resistance with high angular resolution [14]. Subsequently, the intrinsic pinning effect in magnetic and transport properties was also confirmed for other layered cuprates and for iron-based superconductors (see [15] and references therein).

We present the results of a study of the galvanomagnetic properties of epitaxial films of the layered high-temperature superconductor $Nd_{2-x}Ce_xCuO_4/SrTiO_3$ with the c-axis oriented along the short side of the substrate. This configuration is convenient for studying the motion of Josephson vortices in $CuO_2$ layers (along the *c*-axis) taking into account internal pinning effects.

**3.Experimental procedure**

To achieve our objectives, we fabricated $Nd_{2-x}Ce_xCuO_4/SrTiO_3$ ($x$ = 0.145, optimally doped region) epitaxial films synthesized by pulsed laser deposition by A. Ivanov (National Research Nuclear University MEPhI - Moscow Engineering Physics Institute) with $(1\bar{1}0)$ orientation when the *c*-axis is directed along the short side of the substrate [16]. This orientation was chosen deliberately to measure the Hall effect between the $CuO_2$ planes.

The films were fabricated as a Hall bridge (Fig. 1): film thickness 520 nm, width 0.8 mm, length 5.0 mm, lead wire thickness 0.05 mm.

The transverse voltage ($U_y$) was measured at *dc* by reversing the sign of the external magnetic field across the Hall contacts. The electric field was always applied parallel to the plane of the $SrTiO_3$ substrate along the $CuO_2$ layers. The external magnetic field ***B*** was always directed perpendicular to the plane of the $SrTiO_3$ substrate and along the $CuO_2$ layers.

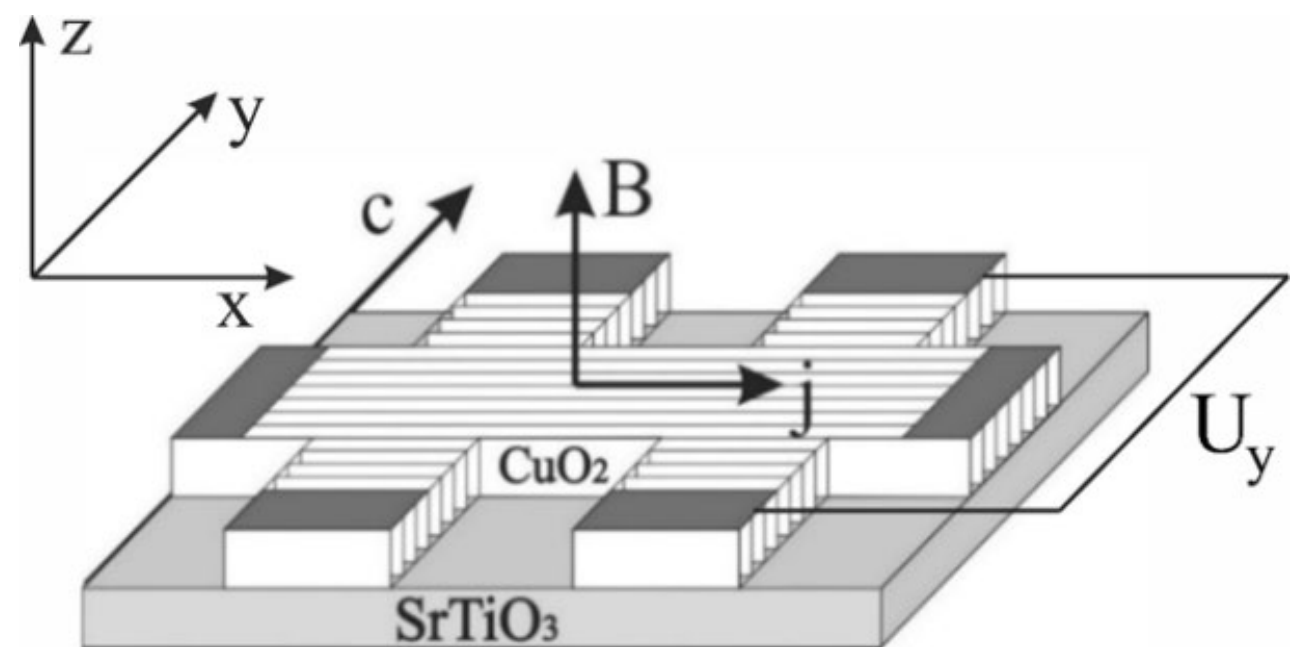

**Fig. 1** Schematic representation of the films studied

The study of magnetic field dependences of transverse voltage with changing current for $Nd_{2-x}Ce_xCuO_4/SrTiO_3$ films was carried out on the original certified setup for measuring galvanomagnetic effects with a solenoid "Oxford Instruments" (Collaborative Access Center (CAC) "Testing Center of Nanotechnology and Advanced Materials" of the Institute of Metal Physics of the Ural Branch of the Russian Academy of Sciences) in magnetic fields up to 9 T at helium temperatures, $T$ = 1.7 K and 4.2 K.

## 4.Experimental results and discussion

The configuration of $Nd_{2-x}Ce_xCuO_4/SrTiO_3$ epitaxial films chosen in this study, with the *c*-axis oriented along the short side of the substrate, is convenient for studying the motion of Josephson vortices across the $CuO_2$ layers (along the *c*-axis), considering intrinsic pinning effects. The external magnetic field in our experiment was directed parallel to the conducting $CuO_2$ planes. In this case, as $B$ increases (at $B>B_{c1}$ = 0.07 T), a system of Josephson vortices forms in the sample.

Figure 2 shows the current-voltage (CV) characteristics $j(U_y)$ in a sample of $Nd_{2-x}Ce_xCuO_4$ with $x$ = 0.145 at temperatures of 1.7 K and 4.2 K, for one of the directions of the external magnetic field, $B_{\parallel}^{-} = (0-(-9))$T, and two directions of the applied electric current.

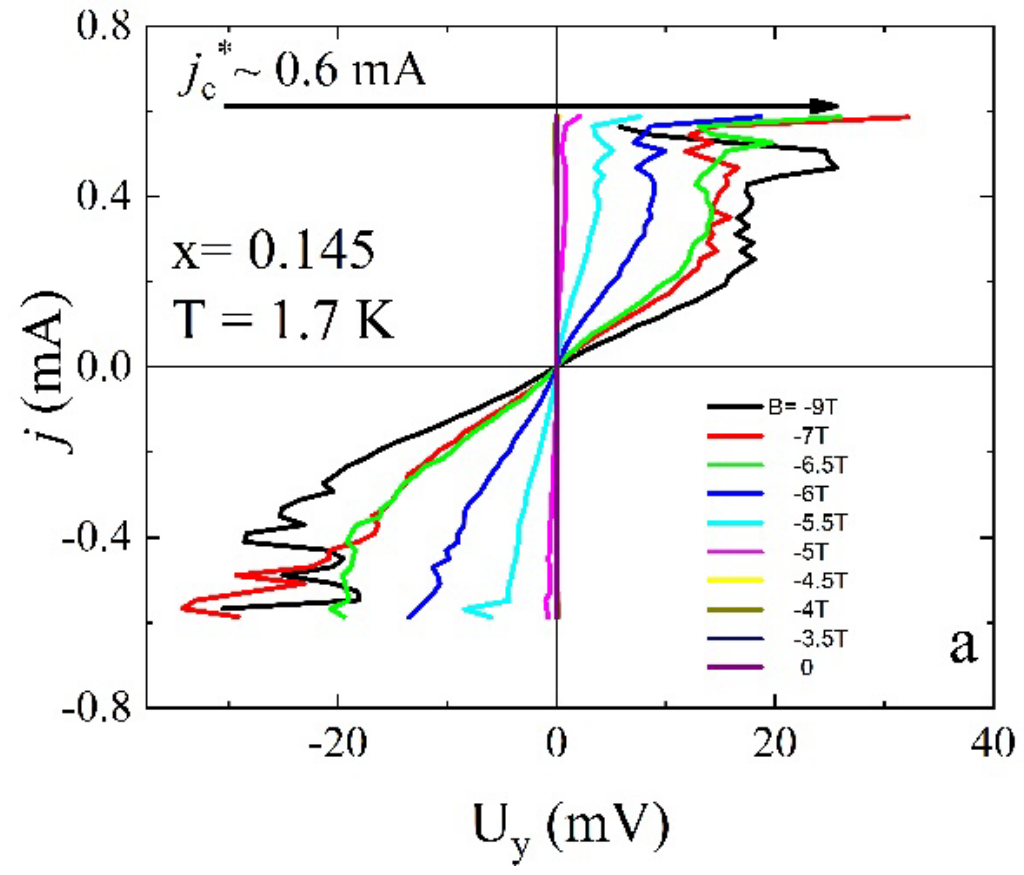

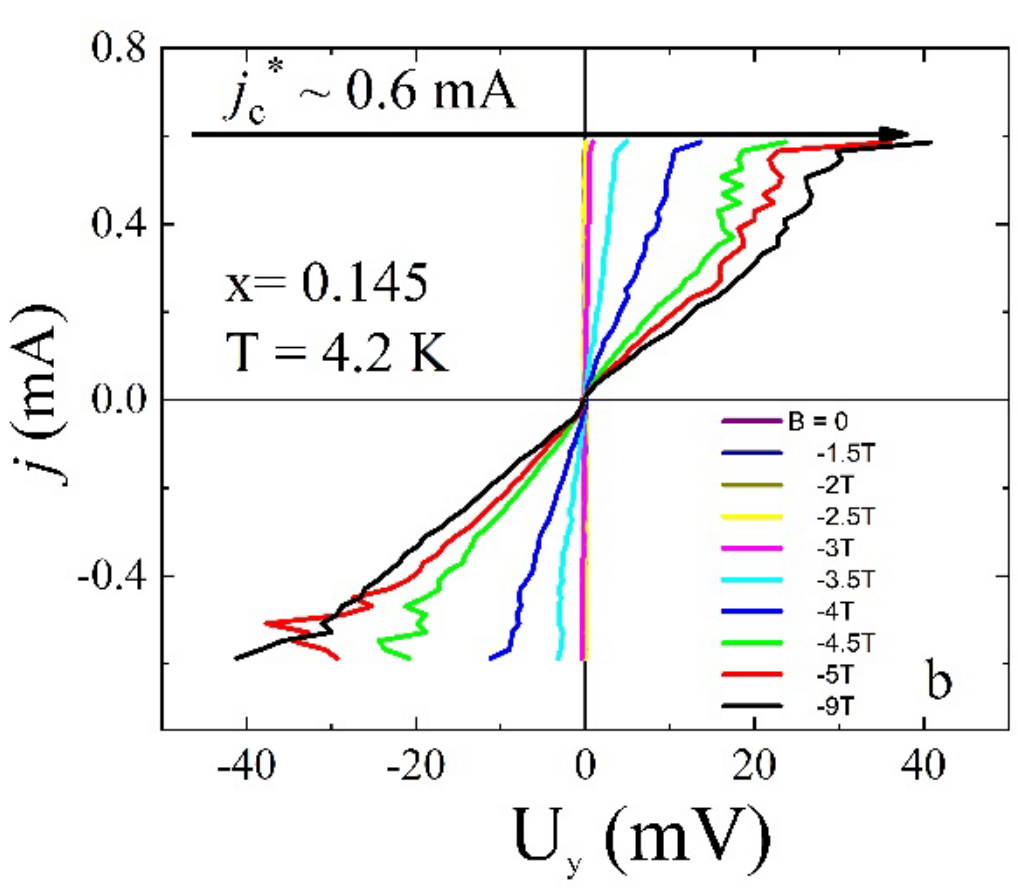

**Fig. 2** CV characteristics of the $Nd_{2-x}Ce_xCuO_4/SrTiO_3$ with $x = 0.145$ at the Hall contacts in a fixed magnetic field ($B = (0 – (-9))$ T) at $T = 1.7$ K (a) and $T = 4.2$ K (b) in the configuration $\boldsymbol{j}||CuO_2$, $\boldsymbol{B}||CuO_2$ and $\boldsymbol{j}\perp\boldsymbol{B}$. The arrows indicate the transition to the main resistive branch at $j_c^* \sim 0.6$ mA

*Critical depinning current of Josephson vortices.*

From Fig. 2a, it is evident that at $T = 1.7$ K in magnetic fields B < 5 T, with electric currents up to $j = 0.6$ mA, the voltage across the Hall contacts is not detected. The finite voltage, corresponding to the transition to a resistive state caused by the motion of the Josephson vortex system (flux-flow resistive state), appears when the critical depinning current, $j_c(B_{\|})$, becomes less than the maximum applied external current. In magnetic field $B = 5$ T, the critical current is $j_c = 580$ μA, and decreases exponentially with further increase in the magnetic field (see Fig. 3).

At $T = 4.2$ K, the depinning magnetic field decreases and is equal to ~ 3T (Fig. 2b). In Fig. 2 a, b, at the maximum applied external current, over the entire range of magnetic fields, we see a tendency toward a transition to the main resistive branch of the system of intrinsic Josephson junctions in the $Nd_{2-x}Ce_xCuO_4/SrTiO_3$ films (critical transition current $j_c^* \sim 0.6$ mA).

Figure 3 shows the values of the critical depinning current of the studied sample as a function of the magnetic field at $T = 1.7$ K and 4.2 K. In a parallel field, $j_c(B)$ changes as $\exp(-B/B_0)$ with $B_0$ in the range of (0.5 – 0.7) T (Fig. 3b) in accordance with the expected behavior of Josephson vortices in the regime of pinning [2. 3].

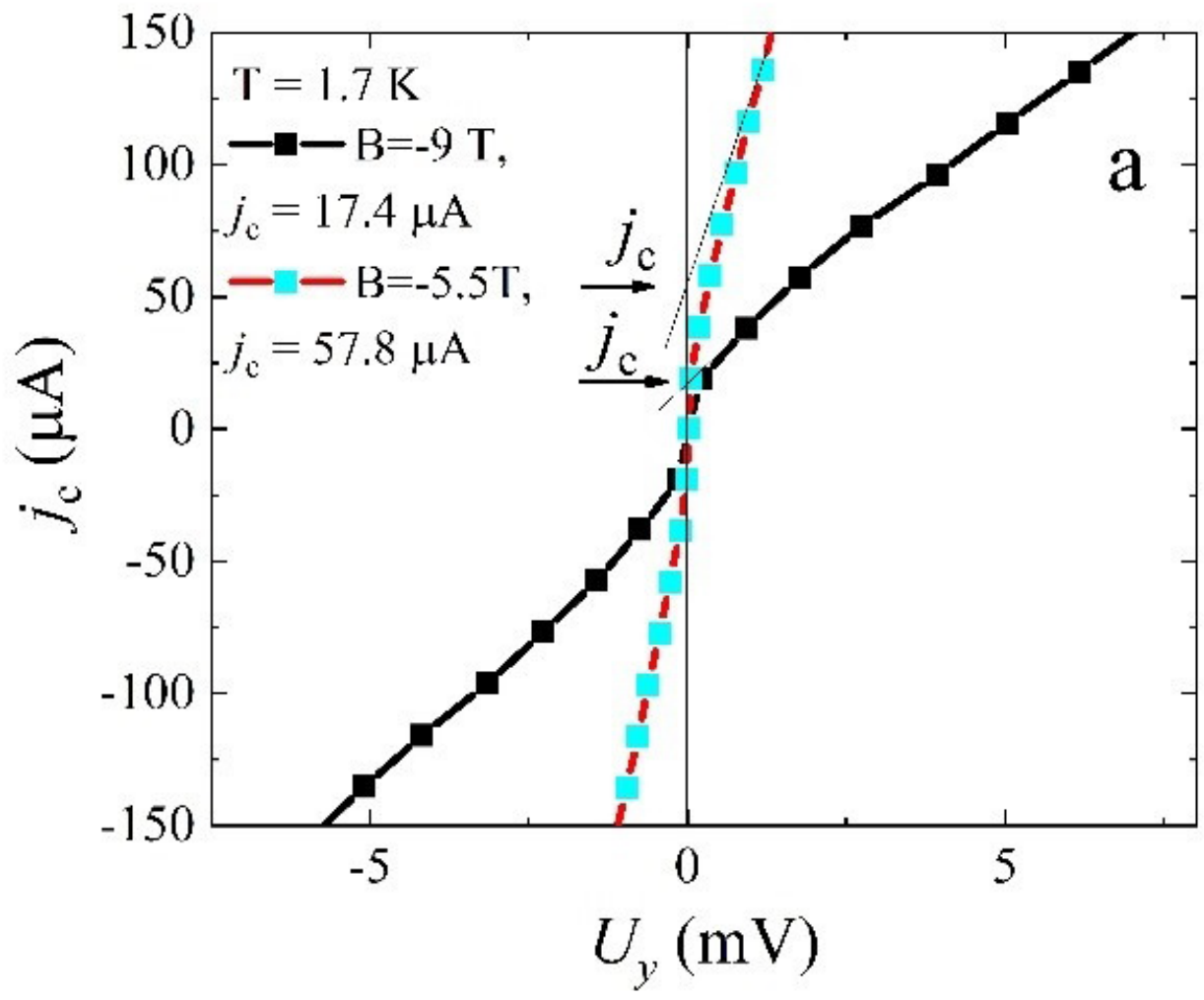

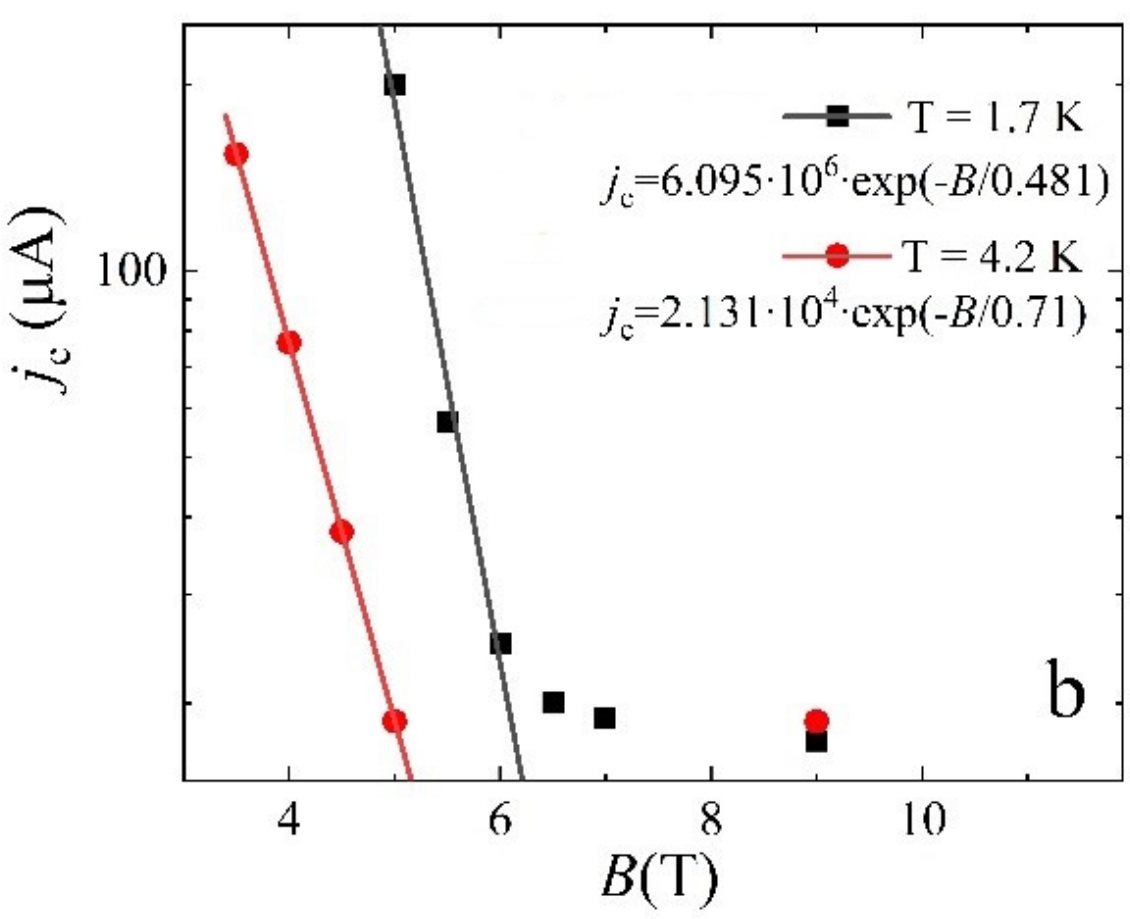

**Fig. 3** (a) CV characteristics at the Hall contacts in a magnetic field of $B$ = - 9 T and -5.5 T at $T$ = 1.7 K, the arrows show the critical current, $j_c$, of the Josephson junctions, (b) dependences $j_c(B)$ in a logarithmic scale: $j_c(B) \sim \exp(-B/B_0)$, $B_0$ = 0.48 T at 1.7 K and $B_0$ = 0.71 T at 4.2 K.

In [2. 3], in $Bi_2Sr_2CaCu_2O_x$ structures, an exponential dependence $I_c(B)$ was observed for Josephson vortices in contrast to the power law dependence for Abrikosov vortices. Thus, in [3] over an appreciable region of the phase diagram the critical current $I_c(B_\perp)$ varies $\sim 1/B_\perp^\mu$ , with $\mu \approx$ 0.8 to 0.9. In a parallel field, $I_c(B_\parallel)$ varies as $\exp[-(B_\parallel/B_0)]$, with $B_0 \sim 2$T, consistent with a model involving pinned Josephson vortices.

*Self-resonance effects (Fiske steps).* In the flux-flow regime, a series of current maxima (steps) appears on the CV characteristics in a constant magnetic field, the amplitude of which increases with increasing magnetic field (see Fig. 2a, b). We associate the maxima on $j(U_y)$ with internal resonances (self-resonances) of the Josephson current frequency with the frequency of Swihart waves—electromagnetic waves generated in Josephson junctions in a magnetic field.

The maximum of current on the CV characteristic reflects the intensity of the resonant interaction of the Josephson current with the electromagnetic field, when the phase velocity of the Josephson current wave $v_0$ approaches the velocity $c_0$ of electromagnetic waves that can propagate in the junction (Swihart waves) [1]. When the condition $v_0 = c_0$ is being carried out, a maximum of current $j$ occurs at a voltage value of:

$$U = \Lambda \frac{c_0}{c} B, \quad (1)$$

where $\Lambda = \lambda_{L1} + \lambda_{L2}$ is the sum of the London penetration depth of two superconductors forming Josephson junction.

In multilayer HTSC crystals, which are essentially a stack of many Josephson junctions, a complex superposition of resonant structures (Fiske steps) arises, which is determined by the multitude of intrinsic Josephson junctions and the variation of the Swihart velocity values from layer to layer [17, 18, 19].

Our observation of a series of current maxima in the $j(U_y)$ dependences indicates the self-resonant features of intrinsic Josephson junctions in the $Nd_{2-x}Ce_xCuO_4$ compound and their manifestation in the form of Fiske steps.

*Depinning of Josephson vortices by magnetic field.*

In this subsection, we will consider the breakdown process for intrinsic pinning (depinning) of Josephson vortices by the magnetic field in the studied $Nd_{2-x}Ce_xCuO_4/SrTiO_3$ films.

From the graphs in Fig. 2 we see that at $T = 4.2$ K the sample is in the superconducting state, $U(j) = 0$ (with the accuracy of our measurements) up to fields $B_c = (3 - 3.5)$ T, which corresponds to the absence of motion of the Josephson vortices due to strong intrinsic pinning. The emergence of a resistive state at $B > B_c$ indicates the onset of motion of the vortex lattice (flux-creep or flux-flow regimes) under the action of the Lorentz force, which, as is known [4,5,12], leads to the appearance of an electric field both along (magnetoresistance) and across (Hall effect) the direction of the transport current.

The experimental CV characteristics $j(U_y)$ for the studied sample at $T$=1.7 K are similar, but the depinning field $B_c = (4.5 - 5)$T (Fig. 2a).

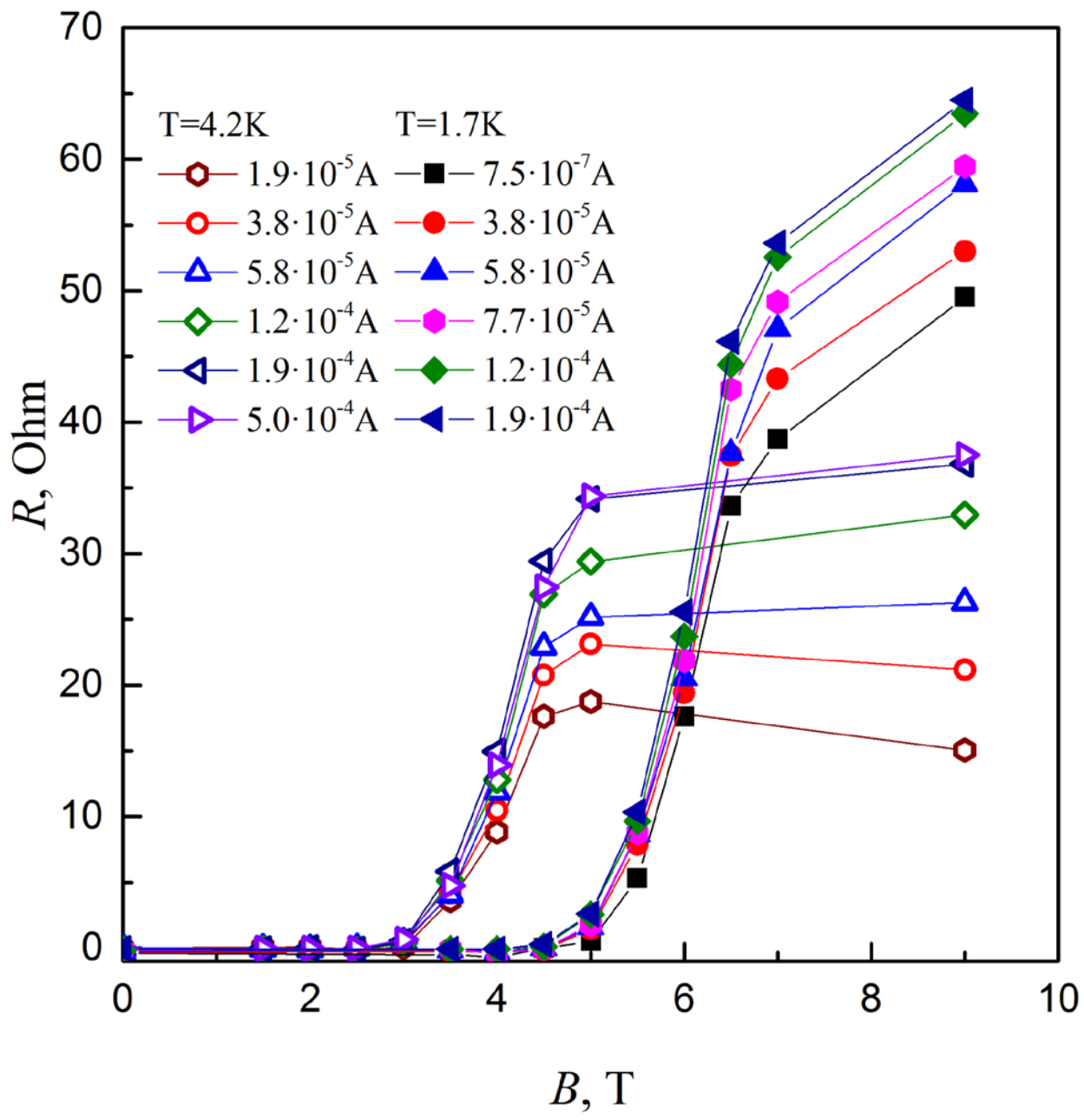

**Fig. 4** Dependences of longitudinal resistance, $R$, on the magnetic field for fixed values of $j$ after averaging the data for $U_y(j, B)$ on the directions of current and field for $T$ = 1.7 K and 4.2 K.

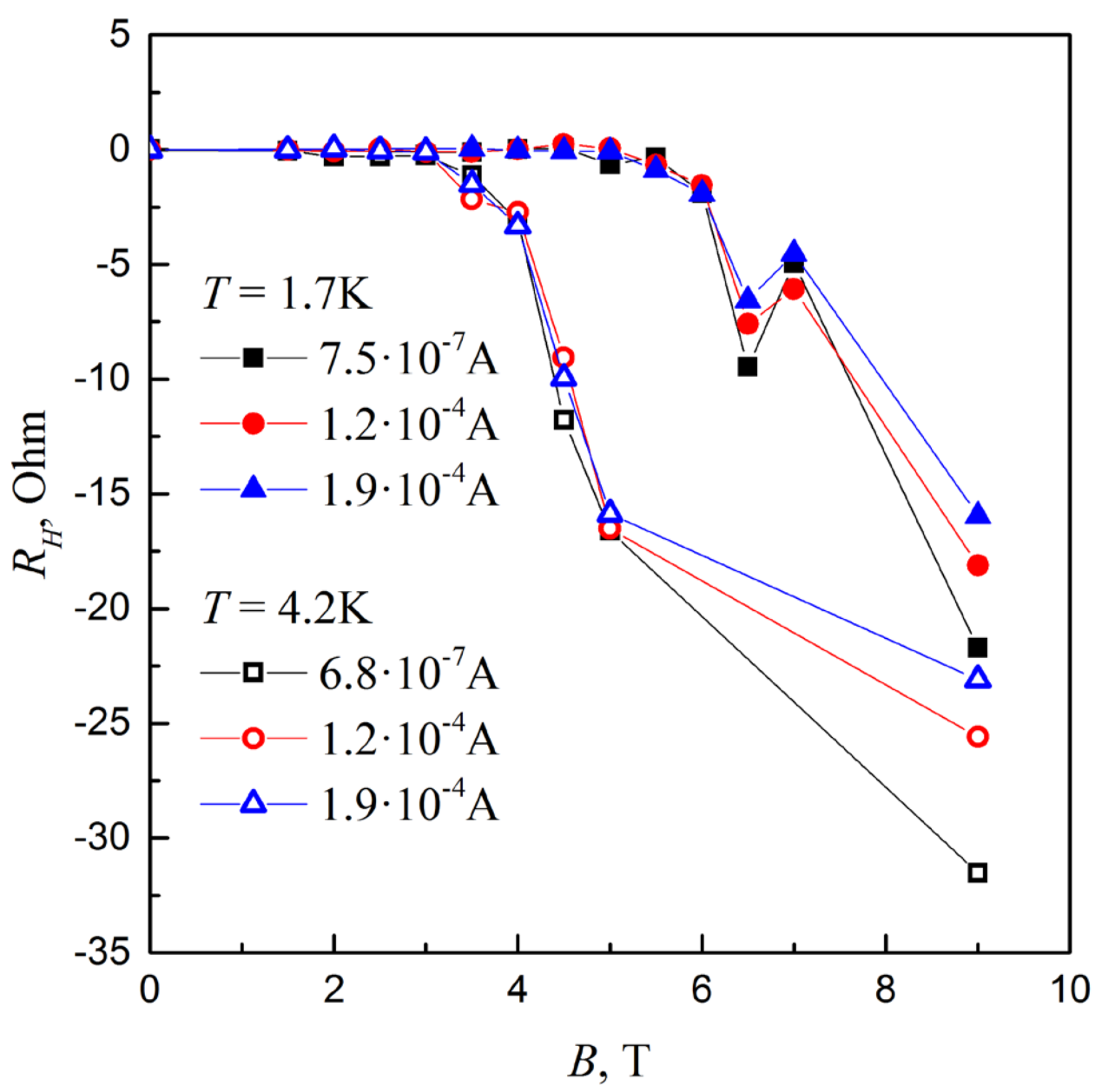

**Fig. 5** Dependences of Hall resistance, $R_H$, on the magnetic field for fixed values of $j$ after averaging the data for $U_y(j, B)$ on the directions of current and field for $T$ = 1.7 K and 4.2 K.

After the standard procedure of averaging over the directions of the external magnetic field, from the experimental data for $U_y(j, B)$, we obtained the dependences of the magnetoresistance, $R\ (B)$ (Fig. 4), and the Hall resistance, $R_H(B)$ (Fig. 5), for fixed values of $j$ at $T$ = 1.7 K and 4.2 K. Here, $R\ (B)$ is the even contribution to the magnetic field, $R_H(B)$is the odd contribution to the magnetic field:

$$R\ (B) = \frac{U_y\ (B) + U_y\ (-B)}{2\ j}, \tag{2}$$

$$R_H(B)\ = \frac{U_y\ (B) - U_y\ (-B)}{2\ j}. \tag{3}$$

We note that in our samples the contribution to $U_y(j, B)$ from the magnetoresistance (symmetrical with respect to $B$) is about 60% of the total voltage, which, as can be shown, is due to the strong anisotropy of the $Nd_{2-x}Ce_xCuO_4$ compound, $\rho_c/\rho_{ab} \approx 400$, where $\rho_c$ and $\rho_{ab}$are resistivity along the $c$-axis and along the $ab$-plane, respectively [20]. Moreover, it has been empirically established that this method is more sensitive than measurements of the CV characteristics of $U_x(j, B)$.

The type of dependencies $R(B)$ and $R_H(B)$ found are in accordance with the concepts of intrinsic pinning with a depinning field $B_c = (4.5 - 5.0)$ T at $T$=1.7 K and $B_c = 3$ T at $T$=4.2 K.

As can be seen from Fig. 4, that at $B > B_c$, our system exhibits a sharp increase in resistance $R(B)$, followed by saturation (at 4.2 K) or a tendency toward saturation (at 1.7 K). The increase in resistance corresponds to the process of carrier energy dissipation due to the onset of Josephson vortex lattice movement.

Thus, in general, the dependence $R(B)$ that we observe looks like this: a pinning region with $R = 0$ at $B < B_c$, then, at $B > B_c$, a steep increase with the "knee" in the $R(B)$ curves at a certain $B(= B_f)$.

The relationships of such kind have been repeatedly observed at $T < T_c$ for HTSC systems with different nature of vortex pinning both for $R(B)$ dependencies at fixed $T$ and for $R(T)$ dependencies at fixed $B$, for Abrikosov vortices (AV) at $B||c$ and/or Josephson vortices (JV) at $B||ab$ (see [21], [22], [23] and also [15] with references therein).

The dissipation due to Lorentz-force driven flux motion has been studied in Y-Ba-Cu-0 single crystals [21], [22], in five groups of high temperature hole doped cuprates [23] as well as for Josephson vortices in iron-based superconductor $FeSe_{1-x}S_x$ with strong intrinsic pinning. In [23], the dissipation increases starting with $Ba_2YCu_3O_7$ ($T_c$ = 88 K), ongoing to $Ba_2YCu_3O_{6.7}$ ($T_c$ = 60 K), $Pb_2Sr_2RCu_3O_8$ ($T_c$ = 50 K), $Bi_2Sr_2CaCu_2O_8$ ($T_c$ = 84 K) and $Tl_2Ba_2CaCu_2O_8$ ($T_c$ = 90 K).

Malozemoff et al. [22] convincingly argued that in these works, in the resistivity of high-$T_c$ superconductors after the depinning, a thermally activated flux-creep mode is first realized and then a free motion of vortices (flux-flow mode) takes place. It is the crossover between flux creep and flux flow that may explain the "knee" on the resistivity curves observed in many works.

We use the formulas of a thermally activated flux-creep model presented in [22] to describe our experimental data for $R\ (B)$ or $R_H(B)$ from Figs. (4) and (5). For analytical treatment of thermal excitation over a potential barrier $U$ we have:

$$R = R_0(U/kT)\exp(-U/kT) \qquad (4)$$

or $R_H = R_H^0(U/kT)\exp(-U/kT)$ (5)

with the pinning potential of the form

$$U(B) = A/B^p. \qquad (6)$$

In Eq. (6) we have added a fitting parameter $p$ in the power law for $U(B)$. Then we get

$$R\ (B) = R_0 \frac{A}{kT}\frac{1}{B^p}\exp\left(-\frac{A}{kT}\frac{1}{B^p}\right) \qquad (7)$$

and a similar formula for $R_H(B)$.

Fig. 6 shows an example of fitting of the observed dependences for magnetoresistance $R\ (B)$ and Hall resistance $R_H(B)$ by Eqs (7) for certain values of temperature and current ($T = 1.7\ K, j = 7.5 \cdot 10^{-7} A$). The found values of the fitting parameters are shown in the field of the figures.

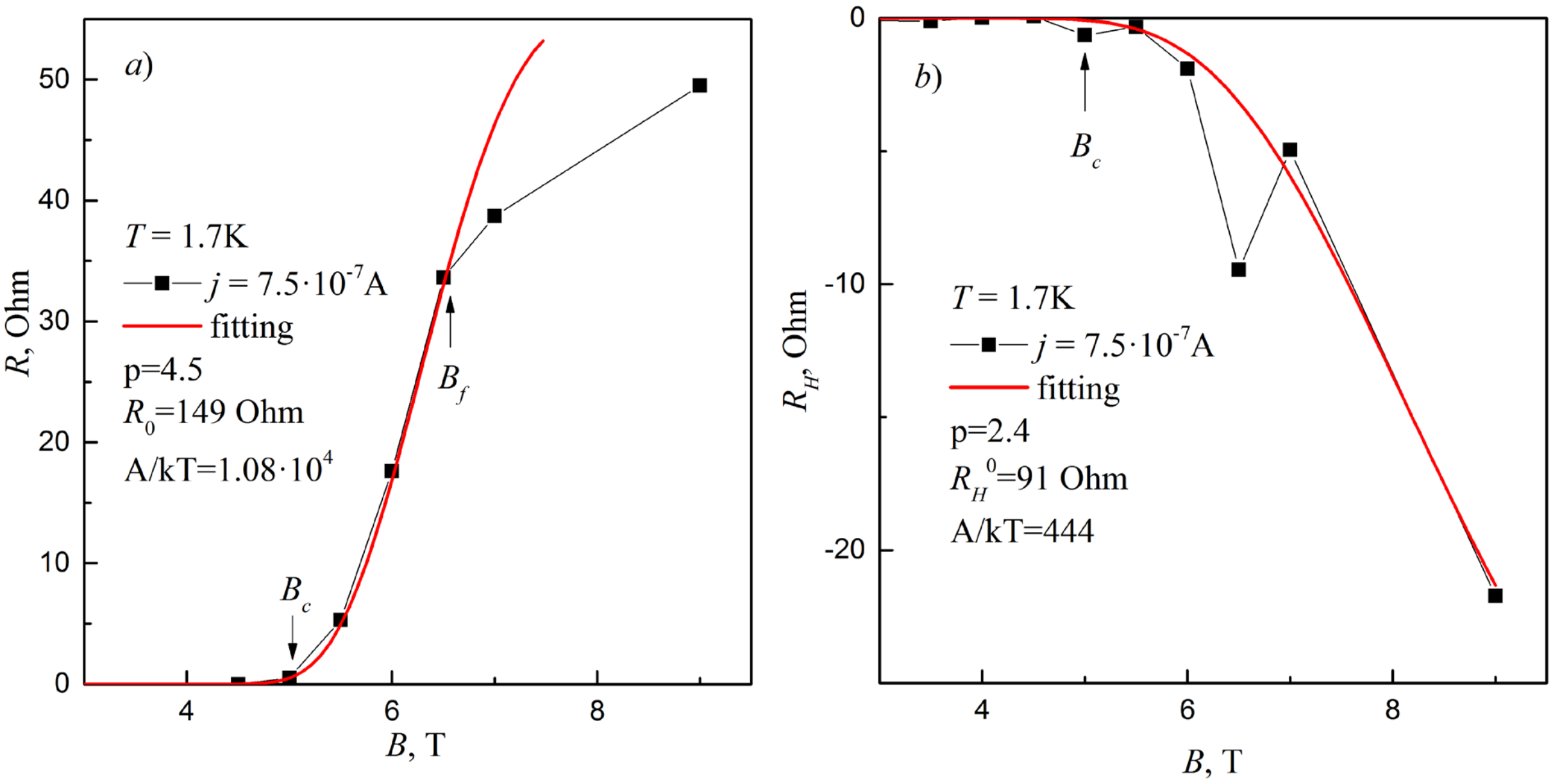

Fig. 6. The results for the description of experimental dependencies (a) of magnetoresistance $R\ (B)$ and (b) Hall resistance $R_H(B)$ with fitting by Eqs (7) at $T = 1.7\ K, j = 7.5 \cdot 10^{-7} A$.

In Fig. 6, we see an adequate description of our experimental data by the expression (7) for the flux creep mode up to crossover magnetic field $B = B_f$, field for transition to the flux flow mode: $B_f = 6.5$ T for $R\ (B)$ at 1.7 K. From Figures 4 and 5 it can be seen that $B_f = 4.5$ T at 4.2 K for $R\ (B)$, and that $B_f = 5$ T at 4.2 K and $B_f = 9$ T at 1.7 K for $R_H(B)$.

A crossover magnetic field, $B = B_f$, is defined by the condition that the effective activation energy for flux motion is of the order of $kT$: $U/kT \approx 1$. Let us note that an extrapolation of exponential dependence of $j_c\ (B)$ to zero (see Fig. 3b) corresponds well to the crossover at $B = B_f$ on the free flow of vortices, which is in agreement with the physical meaning of this crossover.

The best fit of the experimental dependences under the conditions of Fig. 6 was obtained at $p = 4.5$ for $R\ (B)$ and at $p = 2.4$ for $R_H(B)$. The difference in the power-law dependences $U(B)$ for the different resistance tensor components may be due to both physical reasons (different current flow paths for $\rho_{xx}$ and $\rho_{xy}$) and the need to use more sophisticated formulas for systems with intrinsic type of the pinning.

Note that for barriers created by lattice defects under conditions of Abrikosov vortex creep, the theory gives $U_{pin} \sim 1/B$ [21], [22], while the experiment yields $U_{pin} \sim B^{-n}$ with $n = 0.5$ for strong pinning in YBaCuO [24], $n = 1$ and $n = 1.33$ for weak pinning in LaSrCuO [25] and $MgB_2$ [26], respectively.

The behavior of $R\ (B)$ at $B > B_f = 6.5$ T, for 1.7 *K,* can be described by the simplest form of the Bardeen-Stephen flux flow resistivity [27]:

$$R\ (B) = R_n\, B/B_{c2} \qquad (8)$$

with the second critical field $B_{c2} = 9$T and the normal state resistance $R_n(j = 7.5 \cdot 10^{-7} A) \approx 50 Ohm$ (see Figs 4 and 6a).

As for the Hall resistance, to our knowledge, there has not yet been an analytical description for $R_H(B)$ within a thermally activated flux-creep model. From Fig. 5 it can be seen that after depinning, at $B > B_c$, the sign of the Hall effect is negative ($R_H < 0$) in accordance with the sign of the carriers in the electron-doped NdCeCuO system.

In a narrow range of magnetic fields, $B_c < B < B_f$ (thermally activated flux creep mode), the value of $|R_H|$ increases sharply. These are the regions $B \cong (3 - 5)$T for 4.2 *K* and $B \cong (5 - 7)$T for 1.7 *K*. The further observed slowdown in the decline of $R_H(B)$ corresponds to the transition on the flux flow mode.

The features of the behavior of the Hall resistance $R_H(B)$in the mixed and normal states of $Nd_{2-x}Ce_xCuO_4/SrTiO_3$ films ($x = 0.135$, 0.145, and 0.15) for three different types of orientation of the *c*-axis relative to the substrate were discussed in detail by us earlier in [28].

*The behavior of* $B_c$ *vs* $j$.

Let us comment on another unusual property of the depinning process observed in our system. From Figs. 4 and 5, it is evident that the depinning field, $B_c$, is independent on the transport current, $j$, at least in the investigated range of $j$ values. This means that the reduction in the intrinsic pinning barrier cannot be explained solely by the action of the Lorentz force, $\boldsymbol{F_L} = \boldsymbol{j} \times \boldsymbol{B}$. From a physical point of view, it is reasonable to associate this fact with the transformation (evolution) of the Josephson vortex system with increasing magnetic field.

Unlike Abrikosov vortices in standard anisotropic superconductors, Josephson vortices do not have normal cores. The centers of Josephson vortices are located between the layers, and the normal core is replaced by a nonlinear region within which the phase difference between the two central layers varies from 0 to 2π [11, 29,30].

The size of the nonlinear core is determined by the Josephson length $\lambda_J = \gamma d$ along the *ab*-planes (i.e., along the layers) and the length *d* along the *c*-axis (where the Josephson nature of the interlayer current is important). Here, $d$ is the periodicity of the layers, $\gamma$ is the anisotropy of the London penetration depth, $\gamma = \lambda_c/\lambda_{ab}$, where $\lambda_c$ and $\lambda_{ab}$ are the penetration depths for currents perpendicular and parallel to the layers, respectively.

Depending on the magnitude of the applied magnetic field $B$, two distinct regimes exist. The crossover field, $B_{cr}$ separating these two regimes is determined by the expression: $B_{cr} = \Phi_0/2\pi\gamma d^2$, where $\Phi_0 = h/2e$ is the magnetic flux quantum. In weak fields, the Josephson vortices are isolated and form a triangular lattice, strongly stretched along the layer direction. In the dilute limit, $B \ll B_{cr}$, the nonlinear cores of the Josephson vortices are well separated from each other.

When the magnetic field exceeds the crossover field, $B_{cr}$, the centers of the Josephson vortices begin to overlap, and a dense Josephson lattice is gradually formed [11, 29,30]. In fields $B \gg B_{cr}$, the nonlinear cores of the vortices overlap strongly, since the distance between the vortices along the *ab*-plane becomes less than $\lambda_J$ and the interlayer spaces are uniformly filled with vortices in this case. Strong interlayer coupling in the dense limit leads to the movement of the vortex lattice under the influence of the Lorentz force as a whole, overcoming the forces of intrinsic pinning.

The overlap of the Josephson vortex cores is an extremely strong limit; we find $B_{cr} \approx 50$ T for our system ($\gamma = 18, d = 0.6\ nm$). However, we see empirically that already at $B_c \approx 0.1\ B_{cr}$ the density of the vortex lattice is large enough to overcome the internal barriers created by the superconducting $CuO_2$ layers.

## 5. Conclusions

The galvanomagnetic properties of an epitaxial film of layered electron-doped superconductor $Nd_{2-x}Ce_xCuO_4$ ($x = 0.145$) with the *c*-axis oriented along the short side of the $SrTiO_3$ substrate (orientation $(1\overline{1}0)$) were studied. Measurements were performed in crossed electric and magnetic fields parallel to the *ab* planes to investigate the influence of the Lorentz force directed along the *c*-axis on the system of Josephson vortices in strong magnetic fields up to 9 T at helium temperatures, $T = 1.7$ K and 4.2 K.

Our objective is to study the physical processes occurring in a system of Josephson vortices under the influence of a magnetic field when the conditions of strong intrinsic pinning are met. The configuration of $Nd_{2-x}Ce_xCuO_4/SrTiO_3$ epitaxial films chosen in this study, with the *c*-axis oriented along the short side of the substrate, is just convenient for studying the motion of Josephson vortices across the $CuO_2$ layers (along the *c*-axis), considering intrinsic pinning effects.

The evolution of the CV characteristics for the voltage at the Hall contacts, $U_y$, *vs* transport current, $j$, with increasing magnetic field was studied. The analysis of depinning processes both under the action of external current (determination of critical current, $j_c$) and under the influence of magnetic field (estimation of critical field, $B_c$) was carried out.

An exponential dependence of the depinning current $j_c$ on the magnetic field was found in contrast to the power-law dependence $j_c(B)$ for Abrikosov vortices. Further, it has been established that the magnitude of the depinning field, $B_c$, is practically independent of the magnitude of the transport current, which is possible only with a strong influence of the magnetic field on the root cause of intrinsic pinning, namely, on the strength of the coupling between individual vortices in the Josephson lattice. It is known that as the magnetic field increases, a dense lattice of Josephson vortices forms. This lattice begins to move as a single unit under the influence of the Lorentz force, overcoming intrinsic vortex pinning in fields exceeding the critical field $B_c = (3 - 5)$T.

It's inspiring that after the depinning, at $B > B_c$, our system clearly exhibits the flux creep mode and then, with increasing magnetic field, the crossover to flux flow in both resistivity and the Hall effect takes place. This correlates with the observation of such an effect for Abrikosov vortices in YBaCuO crystals and in other hole doped high temperature cuprates (including highly anisotropic Bi- and Tl-systems) as well as for Josephson vortices in iron-based superconductor $FeSe_{1-x}S_x$ with strong intrinsic pinning.

In general, for electron-doped HTSC $Nd_{2-x}Ce_xCuO_4$, experimental confirmation of the theoretically predicted effect of intrinsic pinning of Josephson vortices was obtained, and the regularities of the depinning process under the action of an external current and/or magnetic field were analyzed.

**Acknowledgments**

The work was carried out within the framework of the state assignment of the Ministry of Science and Higher Education of the Russian Federation for the IMP UB RAS.

**Contributions**

T.B. Charikova: conceptualization, writing- reviewing and editing, V.N. Neverov: measurements, M.R. Popov: writing- reviewing and editing, visualization, S.D. Popov: measurements, writing- reviewing and editing, N.G. Shelushinina: theoretical analysis, writing- original draft preparation, A.A. Ivanov: growth of the samples, All authors reviewed the manuscript.